\documentclass{article}%
\usepackage{amssymb}
\usepackage{amsfonts}
\usepackage{amsmath}%
\usepackage{graphicx}
\providecommand{\U}[1]{\protect\rule{.1in}{.1in}}
\newtheorem{theorem}{Theorem}
\newtheorem{acknowledgement}[theorem]{Acknowledgement}

\begin{document}

\title{Exact and approximate analytical solutions of Weiss equation of ferromagnetism
and their experimental relevance}
\author{Victor Barsan$^{\left(  1\right)  }$, Victor Kuncser$^{\left(  2\right)  }$\\IFIN-HH, Str. Reactorului 30$^{\left(  1\right)  }$, and INFM, Str.
Atomistilor 405$^{\left(  2\right)  },$ \\077125 Magurele, Romania}
\maketitle

\begin{abstract}
The recent progress in the theory of generalized Lambert functions makes
possible to solve exactly the Weiss equation of ferromagnetism. However, this
solution is quite inconvenient for practical purposes. Precise approximate
analytical solutions are obtained, giving the temperature dependence of the
spontaneous magnetization, and also the dependence of the magnetization on
both temperature and external magnetic field. The experimental relevance of
these results, mainly for the determination of the Curie temperature, is discussed.

\end{abstract}

\section{Introduction}

The Weiss theory of ferromagnetism is \ an exactly solvable model "that allows
one to study thermodynamic functions in detail, in particular their properties
near the critical temperature"\cite{[Kochmanski]}. The only element which
hampers the possibility of calculating analytically any thermodynamic aspect
is the fact that the equation of state, which is a transcendental equation
involving the magnetization, the temperature and the external magnetic field,
cannot be solved analytically. Recently, this obstacle has been removed, as,
due to work of Mez\"{o}, Baricz \cite{[Mezo]} and Mugnaini \cite{[Mugnaini]},
the equation of state was in principle solved, using generalized Lambert functions.

However, these solutions are quite cumbersome and difficult to handle, at
least in the context of experimental magnetism. This is why it is still
important to have, besides the exact solutions, some simpler analytical
approximations, more convenient in practical situations. In the last decades,
several analytical approximations have been obtained for the Brillouin and
Langevin functions or their inverses, with applications in ferromagnetism
\cite{[Arrott]}, polymers (strong polymer deformation and flow) \cite{[Johal]}%
, \cite{[Kroger]}, \cite{[Jedynak]}, \cite{[Petrosyan]} or solar energy
conversion (daily clearness index) \cite{[Sue]}, \cite{[KeadyL]}. The main
goal of this paper is to give new and precise examples of such analytical approximations.

The structure of this paper is as follows. In Section 2, we shall expose the
main elements of the Weiss theory and the attempts of solving - exactly or
approximately - the equation of state. The exact solutions, in terms of
generalized Lambert functions, are presented. In Section 3, we obtain an
approximate but very precise expression of the magnetization as a function of
temperature, in zero magnetic field, involving elementary functions only. In
Section 4, a similar result is obtained, for the case of a non-zero external
magnetic field. In Section 5, the experimental relevance of these results is
discussed. The last section is devoted to conclusions.

\section{An outline of Weiss theory of ferromagnetism}

The main equation of this theory is the equation of state \cite{[Stanley]},
\cite{[Kittel]}, \cite{[Kuncser1]}, \cite{[Kuncser2]}:%

\begin{equation}
M\left(  T,H\right)  =M_{0}B_{S}\left(  \frac{\overline{\mu}S}{kT}\left(
H+\lambda M\right)  \right)  ,\ \overline{\mu}=g\mu_{B}\label{1}%
\end{equation}
describing the dependence of magnetization $M$ on temperature $T$\ and on
external magnetic field $H$. Also,%

\begin{equation}
M_{0}=N\overline{\mu}S=NSg\mu_{B}.\label{2}%
\end{equation}

So, $M_{0}$ is the maximum value of the spontaneous magnetization.

The parameter $\lambda$ is the molecular field parameter, introduced by Pierre
Weiss in 1907, and $B_{S}$ is the Brillouin function:%

\begin{equation}
B_{S}\left(  x\right)  =\frac{2S+1}{2S}\coth\left(  \frac{2S+1}{2S}x\right)
-\frac{1}{2S}\coth\left(  \frac{1}{2S}x\right) \label{3}%
\end{equation}
If $S=1/2:$%

\begin{equation}
B_{1/2}\left(  x\right)  =\tanh x\label{4}%
\end{equation}

So, if $S=1/2$ and the external field is zero, $H=0,$ eq. (\ref{1}) becomes:%

\begin{equation}
M=M_{0}\tanh\left(  \frac{\overline{\mu}}{2kT}\lambda M\right) \label{5}%
\end{equation}

This equation has a non-zero solution, and, equivalently, the system has a
spontaneous magnetization $M\neq0,$ if the temperature $T$ is under a critical
value, namely under the critical temperature $T_{c},$ given by the following
relation (\cite{[Stanley]}, eqs. (6.18), (6.19)):%

\begin{equation}
T_{c}=\lambda\frac{N\overline{\mu}^{2}}{4k}\label{6}%
\end{equation}

It is convenient to introduce the reduced parameters:%

\begin{equation}
t=\frac{T}{T_{c}},\ h=\frac{\overline{\mu}H}{2kT_{c}},\ \ m=\frac{M}{M_{0}%
}\label{7}%
\end{equation}
as the equation of state (\ref{1}) can be written now in a simpler form:%

\begin{equation}
m\left(  t,h\right)  =\tanh\frac{m\left(  t,h\right)  +h}{t}\label{8}%
\end{equation}
If we write the reduced magnetization in zero magnetic field as%

\begin{equation}
m\left(  t,0\right)  =m\left(  t\right) \label{9}%
\end{equation}
the equation (\ref{1}) becomes:%

\begin{equation}
m\left(  t\right)  =\tanh\frac{m\left(  t\right)  }{t}\label{10}%
\end{equation}
\qquad

Because eq. (\ref{10}) is sometimes called Weiss equation, we shall also adopt
this terminology in the present paper. It is a transcendental equation, as its
solution cannot be written as a finite combination of elementary functions.

There are several attempts of solving, exactly or approximately, this
equation. An interesting approach is due to Siewert and Essing \cite{[Siev]};
it is based on the theory of complex functions, mainly on the theory of
singular integral equations \cite{[Mus]}. It has been systematically applied
by Siewert and his co-workers in order to solve a large number of
transcendental equations, involving hyperbolic (or trigonometric) and
algebraic functions. However, the results obtained in this way are not
explicit and can be hardly used in practical calculations.

Another approach, invented and re-invented by several authors \cite{[Pogosov]}%
, \cite{[VB-PhilMag]}, puts the transcendental equation in a differential form
and writes its solution as a series expansion. If we define the function
$\zeta\left(  t\right)  $\ by\newline%

\begin{equation}
m\left(  t\right)  =t\zeta\left(  t\right) \label{11}%
\end{equation}
the Weiss equation (\ref{10}) takes the form:%

\begin{equation}
\tanh\zeta\left(  t\right)  =t\zeta\left(  t\right) \label{12}%
\end{equation}
We easily find that $\zeta^{\prime}\left(  t\right)  $\ can be expressed in
terms of the function $\zeta\left(  t\right)  $\ and of $t$:%

\begin{equation}
\zeta^{\prime}\left(  t\right)  =\frac{\zeta\left(  t\right)  }{1-t-t^{2}%
\zeta^{2}\left(  t\right)  }\label{13}%
\end{equation}

As any higher order derivative of $\zeta\left(  t\right)  $\ has the same
property, we can write down the series expansion of $\zeta\left(  t\right)
$\ near an arbitrary point $t_{0}$ as:%

\begin{equation}
\zeta\left(  t\right)  =\zeta\left(  t_{0}\right)  +\zeta^{\prime}\left(
t_{0}\right)  \left(  t-t_{0}\right)  +\frac{1}{2}\zeta^{\prime\prime}\left(
t_{0}\right)  \left(  t-t_{0}\right)  ^{2}+...\label{14}%
\end{equation}

This approach has two inconconveniences: (1) the radius of convergence of the
series is quite small, and the expansion must be repeated in the neighborhood
of several points $t_{0}$; (2) we have to check that the series exists in the
neighborhood of $t_{0}$ (this is not the case, for instance, if $t_{0}$ is an
essential singularity of $\zeta\left(  t\right)  $ ).

Recently, a crucial progress was achieved in solving a large class of
transcendental equations, namely using generalized Lambert functions
\cite{[Mezo]}, \cite{[Mugnaini]}, with applications to magnetism and other
branches of physics \cite{[MezoEJP]}, \cite{[BarsanArXiv]}. Let us mention
that the Lambert function $W\left(  a\right)  $ is the solution in $x$\ of the
transcendental equation:%

\begin{equation}
xe^{x}=a.\label{15}%
\end{equation}
In other words, the Lambert function is the inverse of the function appearing
in the left hand side of (\ref{15}). Similarily, the solution in $x $\ of the
transcendental equation:%

\begin{equation}
e^{x}\frac{\left(  x-t_{1}\right)  \left(  x-t_{2}\right)  ...\left(
x-t_{n}\right)  }{\left(  x-s_{1}\right)  \left(  x-s_{2}\right)  ...\left(
x-s_{m}\right)  }=a\ \label{16}%
\end{equation}
is the generalized Lambert function\newline%
\begin{equation}
W\left(  t_{1},...t_{n};s_{1},...s_{m};a\right) \label{17}%
\end{equation}

One of the beneficiaries of the progress made in the study of generalized
Lambert functions is the Weiss equation. Actually, eq. (\ref{12}) takes the form:%

\begin{equation}
e^{2\zeta\left(  t\right)  }\frac{\zeta\left(  t\right)  -\frac{1}{t}}%
{\zeta\left(  t\right)  +\frac{1}{t}}=1\label{18}%
\end{equation}
whose solution can be written as \cite{[BarsanArXiv]}:%

\begin{equation}
m\left(  t\right)  =t\zeta\left(  t\right)  =\frac{t}{2}W\left(  \frac{2}%
{t};-\frac{2}{t};-1\right) \label{19}%
\end{equation}
where $W\left(  \frac{2}{t};-\frac{2}{t};-1\right)  $ is a particular case of
(\ref{17}), namely of $W\left(  t;s;a\right)  $. Clearly, the parameters
$t,\ s$ in $W\left(  t;s;a\right)  $ are particular cases of the parameters
$t_{1},...t_{n},\ s_{1},...s_{m}$ from (\ref{16}), (\ref{17}) and have nothing
to do with any physical quantity, like temperature or spin. At its turn,
$W\left(  t;s;a\right)  $ can be expressed in terms of a simpler, but similar
function, the $r-$Lambert function, $W_{r}.\ W_{r}\left(  a\right)  $ is
defined as the solution of the equation:%

\begin{equation}
xe^{x}+rx=a\label{20}%
\end{equation}
and can be expressed as a series expansion in terms of Mez\"{o} - Baricz
polynomials, $M_{k}^{\left(  n\right)  }\left(  1/\left(  r+1\right)  \right)
$ \cite{[Mezo]}.

So, even if eq. (\ref{19}) provides a direct and exact solution of the Weiss
equation for $h=0$ and $S=1/2$ (e.g. the dependence of the spontaneous
magnetization of a magnetic monodomain system with $S=1/2$ versus
temperature), it is quite unconvenient for practical applications. This is why
a simpler analytical approximation of (\ref{19}) might still be useful.

An interesting approach in finding approximate analytical solutions of
transcendental equations is to approximate the non-algebraic functions
involved in such equations with algebraic ones. In this way, the equation
becomes algebraic \cite{[dAGB]}, \cite{[VB-PhilMag]}. For instance, eq. (45)
of \cite{[dAGB]} suggests the following approximation for the exponential function:%

\begin{equation}
\frac{1}{2}\left(  1-e^{-2z}\right)  \simeq z\frac{1+az}{1+cz}%
,~a=-0.0572,\ c=1.286,\ \ 0<z<5\label{21}%
\end{equation}
Even if such an approximation might give useful results when applied to
certain quantum mechanical problems, it is inappropriate for the Weiss
equation, as it cannot reproduce the critical behavior of the magnetization on
a large enough interval.

In the following section, we shall propose a precise approximate analytical
solution of the Weiss equation, inspired from the standard ways of obtaining
physically convenient solutions of the Schroedinger equation.

\section{A precise approximate analytical solution of Weiss equation}

In order to find a convenient solution of the Weiss equation (\ref{10}) on the
interval $\left(  0,1\right)  $, we shall find out firstly the behavior of
$m\left(  t\right)  $ for $m\left(  t\gtrsim0\right)  $ and $m\left(
t\lesssim1\right)  .$

So, for $t\gtrsim0:$%

\begin{equation}
m\left(  t\right)  =\tanh\frac{m\left(  t\right)  }{t}\simeq1-2\exp\left(
-\frac{2m\left(  t\right)  }{t}\right) \label{22}%
\end{equation}
which can be written as:%

\begin{equation}
2\exp\left(  -\frac{2m\left(  t\right)  }{t}\right)  =1-m\left(  t\right)
\label{23}%
\end{equation}

If we put:%

\begin{equation}
\frac{m}{t}=u,\ m=ut\label{24}%
\end{equation}
eq. (\ref{23}) can be written as:%

\begin{equation}
\exp\left(  -2u\right)  =-\frac{t}{2}\left(  u-\frac{1}{t}\right) \label{25}%
\end{equation}

The most general form of an equation which can be solved using the Lambert
$W\ $function is similar to (\ref{25}):%

\begin{equation}
\exp\left(  -cx\right)  =a\left(  x-r\right) \label{26}%
\end{equation}
and its solution is given by:%

\begin{equation}
x=r+\frac{1}{c}W\left(  \frac{c}{a}\exp\left(  -cr\right)  \right) \label{27}%
\end{equation}
So, comparing (\ref{25}) and (\ref{26}), we have to identify:%

\begin{equation}
c=2,\ a=-\frac{t}{2},\ r=\frac{1}{t}\label{28}%
\end{equation}
and we find the solution of eq. (\ref{25}) as:%

\begin{equation}
u=\frac{1}{t}+\frac{1}{2}W\left(  -\frac{4}{t}\exp\left(  -\frac{2}{t}\right)
\right) \label{29}%
\end{equation}

Consequently:%

\begin{equation}
m=ut=1+\frac{t}{2}W\left(  -\frac{4}{t}\exp\left(  -\frac{2}{t}\right)
\right) \label{30}%
\end{equation}
For $t\gtrsim0$:%

\begin{equation}
W\left(  -\frac{4}{t}\exp\left(  -\frac{2}{t}\right)  \right)  =-\frac{4}%
{t}\exp\left(  -\frac{2}{t}\right)  +...\label{31}%
\end{equation}
where the higher order terms in the expression of $W$ can be neglected
\cite{[Corless]}. Hence, we get:%

\begin{equation}
m\left(  t\gtrsim0\right)  \simeq1-2\exp\left(  -\frac{2}{t}\right) \label{32}%
\end{equation}

So, $m\left(  t\right)  $ cannot be written as a series expansion near the
origin, as $t=0$ is an essential singular point for this function. This
mathematical aspect is unimportant for the physicist, as the correct low
temperature dependence of the magnetization is not exponential, but
power-like; it obeys the well-known "$3/2$ Bloch law" (see also the eqs. (10),
(11) in Ch. 15 of \cite{[Kittel]}, where the previous result, (\ref{32}) of
this preprint, was obtained under the implicit assumption that $m\left(
0\right)  $ is finite).

Near the critical temperature $t_{c}=1,$ the magnetization behaves as%

\begin{equation}
m\left(  t\lesssim1\right)  \sim\sqrt{1-t}\label{33}%
\end{equation}

So, for the whole interval $\left(  0,1\right)  ,$\ it seems convenient to
write the magnetization in the form:%

\begin{equation}
m\left(  t\right)  =\left(  1-2\exp\left(  -\frac{2}{t}\right)  \right)
P_{1/2}\left(  t\right)  \sqrt{1-t}\label{34}%
\end{equation}
where $P\left(  t\right)  $ can be approximate, as precisely as we want, with
a polynomial. As%

\begin{equation}
m\left(  0\right)  =1\rightarrow P_{1/2}\left(  0\right)  =1\label{35}%
\end{equation}
we can presume for $P\left(  t\right)  $\ the following form:%

\begin{equation}
P_{1/2}\left(  t\right)  =\left(  1+a_{1}t+a_{2}t^{2}+a_{3}t^{3}+a_{4}%
t^{4}+a_{5}t^{5}+a_{6}t^{6}+a_{7}t^{7}\right)  =\label{36}%
\end{equation}

\[
=\frac{m\left(  t\right)  }{\left(  1-2\exp\left(  -\frac{2}{t}\right)
\right)  \sqrt{1-t}}%
\]

The coefficients $a_{1},\ ...\ a_{7}$ can be obtained by solving a linear
system of 7 equations, obtained for 7 couples of values $\left(
t_{0},\ m\left(  t_{0}\right)  \right)  .$ The numerical value of $m\left(
t_{0}\right)  $ for a given $t_{0}$ can be obtained solving numerically the
equation: \ %

\begin{equation}
m\left(  t_{0}\right)  =\tanh\frac{m\left(  t_{0}\right)  }{t_{0}}\ \label{37}%
\end{equation}
using (for instance) the FindRoot command of Mathematica. The aforementioned
pairs of values $\left(  t_{0},m\left(  t_{0}\right)  \right)  $ might be
those included in the following list:%

\[
\]

\{0.2 -%
%TCIMACRO{\TEXTsymbol{>} }%
%BeginExpansion
$>$
%EndExpansion
0.9999091217152326`\}

\{0.4 -%
%TCIMACRO{\TEXTsymbol{>} }%
%BeginExpansion
$>$
%EndExpansion
0.9856238716346567`\}

\{0.6 -%
%TCIMACRO{\TEXTsymbol{>} }%
%BeginExpansion
$>$
%EndExpansion
0.9073323166453099`\}

\{0.8 -%
%TCIMACRO{\TEXTsymbol{>} }%
%BeginExpansion
$>$
%EndExpansion
0.7104117834878704`\}

\{0.85 -%
%TCIMACRO{\TEXTsymbol{>} }%
%BeginExpansion
$>$
%EndExpansion
0.6295014763911393`\}

\{0.9 -%
%TCIMACRO{\TEXTsymbol{>} }%
%BeginExpansion
$>$
%EndExpansion
0.525429512658009`\}

\{0.95 -%
%TCIMACRO{\TEXTsymbol{>} }%
%BeginExpansion
$>$
%EndExpansion
0.37948520667808994`\}%

\[
\]

Solving the system of 7 linear equations generated by eq. (\ref{36}) applied
to each from the 7 above pairs, the polynomial $P\left(  t\right)  $\ can be
expressed as:%

\begin{equation}
P_{1/2}\left(  t\right)  =1+0.606\,83t-0.904\,80t^{2}+5.\,\allowbreak
953\,1t^{3}-11.\,\allowbreak705t^{4}+\label{38}%
\end{equation}

\[
+13.\,\allowbreak950t^{5}-8.\,\allowbreak817\,4t^{6}+2.\,\allowbreak
292\,8t^{7}%
\]

Finally, the reduced magnetization is:%

\begin{equation}
m\left(  t\right)  =\left(  1-2\exp\left(  -\frac{2}{t}\right)  \right)
\sqrt{1-t}\cdot\label{39}%
\end{equation}

\[
\cdot\left(  1+0.606\,83t-0.904\,80t^{2}+5.\,\allowbreak953\,1t^{3}%
-11.\,\allowbreak705t^{4}+13.\,\allowbreak950t^{5}-\right.
\]

\[
\left.  -8.\,\allowbreak817\,4t^{6}+2.\,\allowbreak292\,8t^{7}\right)
\]

Let us remind ourselves that, looking for the form (\ref{34}) of the solution
of eq. (\ref{10}), we applied actually an approach systematically used in
order to find a physically acceptable solution of the Scroedinger equation. A
wellknown example of such an approach might be the following: when we are
solving the Schroedinger equation in a Coulombian field, we are looking for a
solution which "behaves well" near origin and at the infinity. It is, in fact,
similar to what we have done here, isolating the behaviour of $m\left(
t\right)  $ near $t=0$ and $t=1.$

The plot of the reduced magnetization $m\left(  t\right)  $ given by eq.
(\ref{39}) is given in Fig. 1.

\begin{figure}
\begin{center}
\includegraphics[width=\textwidth]{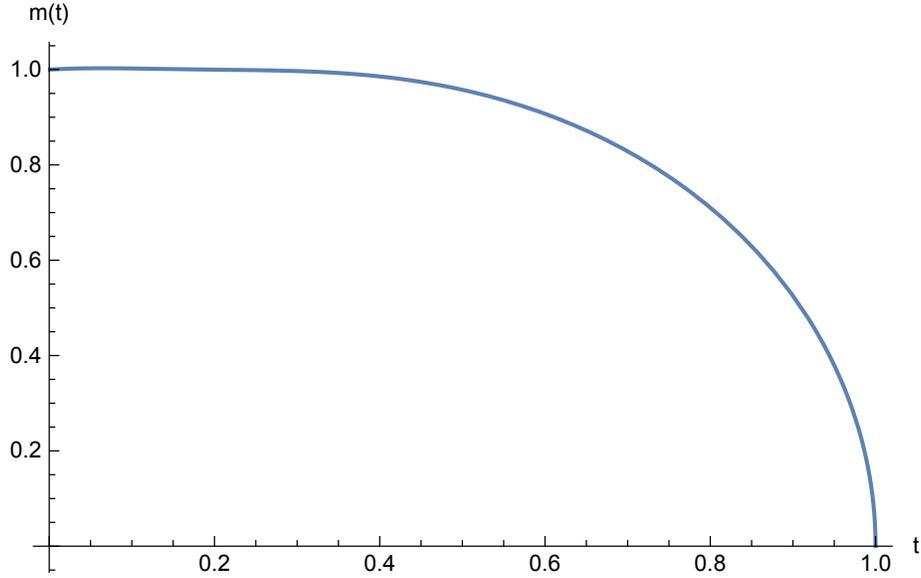}
\end{center}
\caption{The plot of m(t) according to eq. (39).}
\end{figure}

If $m\left(  t\right)  $ satifies (with a good approximation) the equation:%

\begin{equation}
m\left(  t\right)  =\tanh\frac{m\left(  t\right)  }{t}\label{40}%
\end{equation}
it will satisfy (with an equally good approximation) the equation:%

\begin{equation}
m\left(  t\right)  \coth\frac{m\left(  t\right)  }{t}=1\label{41}%
\end{equation}

The function $f\left(  t\right)  =m\left(  t\right)  \coth\left(  m\left(
t\right)  /t\right)  ,$ with $m\left(  t\right)  $given by eq. (\ref{39}) and
the function $g\left(  t\right)  =1$ are plotted in Fig. 2, for comparison.

\bigskip

\begin{figure}
\begin{center}
\includegraphics[width=\textwidth]{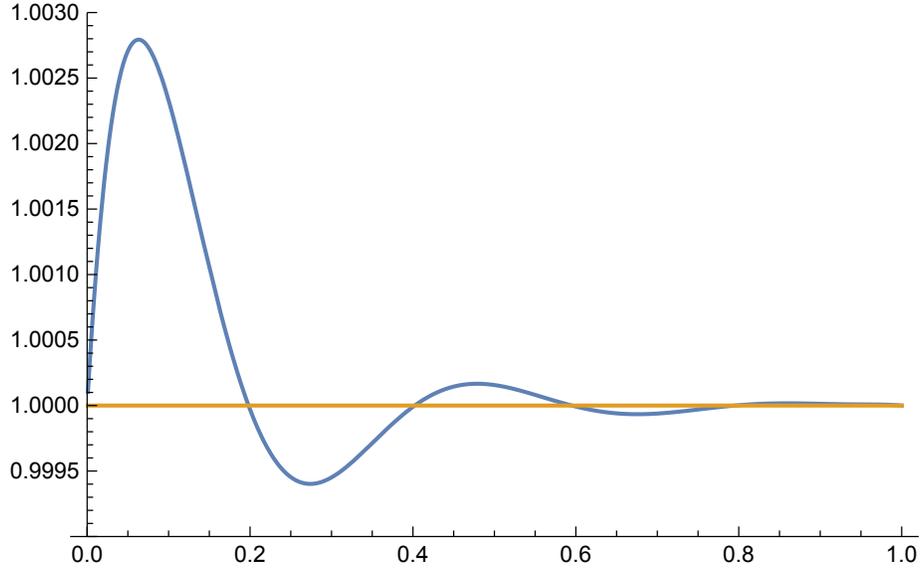}
\end{center}
\caption{The plot of the function from the l.h.s. of eq. (41) and of the constant function y=1.}
\end{figure}

It is easy to see that the approximation of $f\left(  t\right)  $ by $g\left(
t\right)  $\ is excellent. The largest relative deviation, less than
$3\cdot10^{-3}$, corresponds to a region of little physical interest, being
far away from the critical point of the transition temperature, where the
superposition with experimental values should be the most relevant.

In applications to experimental physics involving temperatures of hundread of
Kelvins, the exponential factor in the expression of magnetization plays no
role and instead of (\ref{34}) we can look for a solution of the Weiss
equation (\ref{10}) of the form:%

\begin{equation}
m\left(  t\right)  =Q_{1/2}\left(  t\right)  \sqrt{1-t}\label{42}%
\end{equation}
where $Q_{1/2}\left(  t\right)  $ is a polynomial (of the same degree as
$P_{1/2}\left(  t\right)  $).

Applying the same approach, outlined after eq. (\ref{36}), we get the
following expression for $Q_{1/2}\left(  t\right)  :$%

\begin{equation}
Q_{1/2}\left(  t\right)  =1+0.48248t+0.377693t^{2}+1.66343t^{3}-\label{43}%
\end{equation}

\[
-5.58315t^{4}+6.87009t^{5}-3.99752t^{6}+0.91907t^{7}%
\]

The precision of this simpler expression is still higher then the previous
one: the error is less than $10^{-3},$ for $t\simeq0.1.$

Even more interesting is the fact that this approach can be applied to any
Brillouin function $B_{S}$ defined in (\ref{3}), not only for $S=1/2.$ For an
arbitrary $S,$ the Weiss equation in zero magnetic field is given by the formula:%

\begin{equation}
m(t)=B_{S}\left(  \frac{m(t)}{t}\frac{3S}{S+1}\right) \label{44}%
\end{equation}
and the reduced magnetization can be approximated by:%

\begin{equation}
m\left(  t\right)  =P_{S}\left(  t\right)  \sqrt{1-t}\label{45}%
\end{equation}

For instance%

\begin{equation}
Q_{1}\left(  t\right)  =(1+0.410582t+1.40536t^{2}-3.36245t^{3}+4.34284t^{4}%
-\label{46}%
\end{equation}

\[
-3.28348t^{5}+1.35735t^{6}-0.237194t^{7}%
\]

\begin{equation}
Q_{3/2}\left(  t\right)  =(1+0.410721t+1.65224t^{2}-5.60792t^{3}%
+9.67475t^{4}-\label{47}%
\end{equation}

\[
-9.30572t^{5}+4.74404t^{6}-1.00258t^{7}%
\]

\begin{equation}
Q_{2}\left(  t\right)  =(1+0.453772t+1.34346t^{2}-5.51081t^{3}+10.5506t^{4}%
-\label{48}%
\end{equation}

\[
-10.8288t^{5}+5.77001t^{6}-1.25923t^{7}%
\]

\begin{equation}
Q_{5/2}\left(  t\right)  =(1+0.508037t+0.813671t^{2}-4.26886t^{3}%
+9.01576t^{4}-\label{49}%
\end{equation}

\[
-9.75394t^{5}+5.36827t^{6}-1.19731t^{7}%
\]
with a maximum relative deviation smaller than $1.5\cdot10^{-3},$ also for
$t\simeq0.1;$ for larger values of $t$, the approximation is much better.

\section{The magnetization in the presence of an external magnetic field}

The results of \cite{[Mezo]} can be easily used in order to obtain an explicit
solution of the more general equation of state (\ref{8}), written in reduced
parameters. Actually, according to \cite{[BarsanArXiv]}, this explicit
solution is:%

\begin{equation}
m\left(  t,h\right)  =\frac{t}{2}W\left(  2h+\frac{2}{t};2h-\frac{2}%
{t};-1\right)  -h\ \label{50}%
\end{equation}

For reasons explained in the previous section, this solution is quite
inconvenient for practical purposes. The main goal of the present section is
to obtain an alternative expression for $m\left(  t,h\right)  $, using simpler
functions - actually, a series expansion in powers of $h$, whose coefficients
are known, being expressed as elementary functions of $t.$

With:%

\begin{equation}
\zeta\left(  t,h\right)  =\frac{m\left(  t,h\right)  }{t},\ m\left(
t,h\right)  =t\zeta\left(  t,h\right) \label{51}%
\end{equation}
the equation of state (\ref{8}) becomes:%

\begin{equation}
t\zeta\left(  t,h\right)  =\tanh\left(  \zeta\left(  t,h\right)  +\frac{h}%
{t}\right) \label{52}%
\end{equation}

A closed form of $m\left(  t,0\right)  $ was obtained in the previous section,
eq. (\ref{34}), so an accurate analytic approximation of the function
$\zeta\left(  t,0\right)  $ is known, at least down to enough low $t$.
Differentiating (\ref{52}) with respect to $h,$ and putting%

\begin{equation}
\zeta\left(  t,h\right)  +\frac{h}{t}=\xi\label{53}%
\end{equation}
we get:%

\[
t\frac{\partial\zeta\left(  t,h\right)  }{\partial h}=\left(  1-t^{2}\zeta
^{2}\left(  t,h\right)  \right)  \left(  \frac{\partial\zeta\left(
t,h\right)  }{\partial h}+\frac{1}{t}\right)
\]
and:%

\begin{equation}
\frac{\partial\zeta\left(  t,h\right)  }{\partial h}=\frac{1}{t}\frac{\left(
1-t^{2}\zeta^{2}\left(  t,h\right)  \right)  }{\left(  -1+t+t^{2}\zeta
^{2}\left(  t,h\right)  \right)  }\label{54}%
\end{equation}
Consequently:%

\begin{equation}
\left.  \frac{\partial\zeta\left(  t,h\right)  }{\partial h}\right\vert
_{h=0}=\frac{1}{t}\frac{\left(  1-t^{2}\zeta^{2}\left(  t,0\right)  \right)
}{\left(  -1+t+t^{2}\zeta^{2}\left(  t,0\right)  \right)  }\label{55}%
\end{equation}

So, the value of the previous expression is known, as the function
$\zeta\left(  t,0\right)  $ is known, e.g. via $m(t,0)$ provided by
(\ref{39}). This property of the derivative remains valid in any order;
consequently, we can write%

\begin{equation}
\zeta\left(  t,h\right)  =\zeta\left(  t,0\right)  +\left.  \frac
{\partial\zeta\left(  t,h\right)  }{\partial h}\right\vert _{h=0}h+\frac{1}%
{2}\left.  \frac{\partial^{2}\zeta\left(  t,h\right)  }{\partial h^{2}%
}\right\vert _{h=0}h^{2}+...\label{56}%
\end{equation}
In this way, we obtain a series expansion for $m\left(  t,h\right)  $,
according to (\ref{51}), valid for small values of $h.$ It should have in
mind, however, that we did not investigate the convergence of this series.

\section{Discussion on experimental implications}

\qquad From the experimental point of view, a direct expression of the
magnetization versus temperature and field might be of high interest for
getting material dependent parameters of ferromagnetic materials (magnetic
moment and spontaneous magnetization, Curie temperature, molecular field
constant and exchange integral, etc.) through a suitable fitting of the
experimental data. It is worth to mention that case of ferrimagnetic (or even
no completely compensated antiferomagnets, e.g. due to
structural/interfacial/surfacial defects) can be treated in a similar manner
by a simple association of a finite magnetic moment to a formula unit. Some
real (and very often defect) magnetic systems could present an enough high
magnetic disorder temperature, in order to be completely analyzed within the
temperature range of sensitive magnetometers (e.g. SQUIDS). Mainly these are
the cases where a complete estimation of the material parameters can become
effective by fitting the experimental data obtained over a narrower range of
field and temperature via the above discussed theoretical expressions.

A proof of concept on how this procedure can be experimentally implemented for
magnetization measurements versus temperature will be provided for one of the
above analyzed cases, corresponding to a solid state system with $S=1/2$ (a
quenching of the orbital magnetic moment is assumed). To note that the local
magnetic moment of the magnetic entity (atom, molecule, formula unit) can be
verified from the saturation magnetization at very low temperatures which also
approaches the spontaneous magnetization $M_{0}$. Hence, the experimental
function $m(t,h)$ can be obtained if $T,H$ and $T_{c} $ are known. The first
two are experimental parameters, whereas the last one can be obtained via
specific magnetization measurements versus temperature, e.g. as the
temperature where the spontaneous magnetization drops to zero. Very often,
such measurements have to be done at temperatures much higher than the room
temperature, involving therefore additional higher temperature options for
usual magnetometers. Instead, the Curie temperature can be obtained by fitting
the experimental dependence $M(T)$, as collected under a low enough applied
magnetic field, over a much reduced range of temperatures. The aforementioned
general relation (\ref{51}), written under the form $m(t,h)=t\zeta\left(
t,h\right)  $ should be used in this respect.

However, in order to achieve this task, a series of experimental conditions
should be fulfilled. As a first condition, the system might respect the
overall Brillouin dependence of magnetization versus temperature, that is,
size or anisotropy related magnetization relaxation phenomena (e.g.
superparamagnetic behavior) should be avoided. A second condition is related
to the removal of magnetic domains in the system, condition which might be
fulfilled by measuring the magnetization versus temperature under convenient
constant applied magnetic fields. In the case of quite soft magnetic materials
the measurements can be done in low enough magnetic fields (e.g. of magnetic
induction as low as $0.1T$), in order to keep only the term proportional to
$h$ in relation (\ref{51}). Accordingly, the following experimental steps are
proposed. The $M(H)$ curve at the lowest temperature (e.g. $2K$) should be
initially obtained for estimating the saturation field and the saturation
magnetization $M_{0}$. Then, the dependence $M(T)$ can be obtained under an
applied field higher than the saturation field and up to the maximum
temperature of the device (e.g. 400K for a SQUID). According to Figs. 1 and 2,
the most reliable values for $T_{c}$ might be obtained if the maximum
experimental temperature will overpass at least $60\%$ from $T_{c}$. That is,
for a maximum measuring temperature of $400K$, a Curie temperature of about
$700K$ may be determined with acceptable accuracy.

The procedure of determining the transition temperature is the following. The
experimental ratio $M(T,H)/M_{0}$ will be represented versus $t=T/T_{c}$ for a
set of $T_{c}$ values (e.g. $500K,\ 550K,\ 600K,\ 750K,$\newline$800K$). In
this way, the experimental representations of $m(t,h)$ versus t will be
obtained for different values of $T_{c}$. The theoretical representations of
$m(t,h)$ via the relation $m(t,h)=t\zeta\left(  t,h\right)  $ with
$\zeta\left(  t,h\right)  $ given by eq. (\ref{51}) will be obtained for the
same set of $T_{c}$ values. In a first approximation, $T_{c} $ is estimated as
that temperature value in the considered set corresponding to the best
superposition between the experimental and the theoretical $m(t,h)$ curves.
The procedure can be repeated for a new set of $T_{c}$ values (with a better
temperature discretization, e.g. of $10K$) around the first approximation
until the desired precision of $T_{c}$ is obtained.

It is also to mention that actually there are experimental techniques
providing parameters proportional to the spontaneous magnetization. Among
them, the local techniques of $^{57}Fe$ Mossbauer spectroscopy is extremely
powerful in following the temperature variation of the $Fe$ magnetic moment on
each $Fe$ phase (the hyperfine magnetic field, $B_{hf}$ as specific parameter
is proportional in certain circumstances to the magnitude of the $Fe$ magnetic
moment and hence with the spontaneous magnetization attributed to that $Fe$
phase). Being a local techniques seeing directly the magnitude of the magnetic
moment, it does not involve any applied magnetic field and therefore, the
simplest equation (\ref{51}) can be used as reflecting the experimental
temperature dependence of the ratio $m(t)=B_{hf}(T)/B_{hf}(T=0). $

\section{Conclusions}

A proof of concept for direct series-expansion solutions of the Weiss equation
of ferromagnetism is provided. Even the concept is exemplified in case of
non-zero magnetic field for a magnetic system with $S=1/2$, it works in
principle also for systems with enhanced $S$ values, with the mention that in
that case the Brillouin function will replace the $\tanh$ function, with
direct implications on its derivatives. Also the spontaneous magnetization and
Curie temperataure will depend specifically on $S$, which might become an
additional fitting parameter in the final expression of magnetization. The
general solution of the reduced magnetization in applied field is always
expressed as function of the temperature dependent solution of the spontaneous
magnetization in the absence of the applied field. Specific procedures to use
the developed expressions in case of magnetization or Mossbauer spectroscopy
measurements are also mentioned.

\begin{acknowledgement}

\end{acknowledgement}

The authors acknowledge the financial support of IFIN-HH - ANCSI project PN 16
42 01 01/2016 and to the IFIN-HH - JINR Dubna grant 04-4-1121-2015/17 (VB) and
PN16-480102 (VK).

\bigskip

\bigskip

\bigskip

\end{document}